\documentstyle[prl,aps,epsf,multicol]{revtex}
\begin{document}
\begin{multicols}{2}
\noindent 

{\bf Oreg and Finkel'stein Reply:} In the preceding Comment \cite{Comment}
Fabrizio and Gogolin (FG) dispute our result of the enhancement of the
tunneling density of states (DOS) in a Tomonaga--Luttinger (TL) liquid at
the location of a backward scattering defect \cite{OFa}. They state that (i)
the anticommutativity of the fermion operators of the left and right moving
electrons was not considered properly; and that (ii) the
tunneling DOS\ for this problem is given by $\rho \left( \omega \right) \sim
\omega ^{1/g-1}$ which is the tunneling DOS\ at the end of a TL\ wire. We
disagree with both points. Moreover, we show below that the result of the
Letter can be reproduced following the Comment when its calculations are
performed correctly.

(i) There are different ways to get the correct commutation relations of the
fermion operators in the bosonization technique. We have employed the
variant with a proper regularization of the commutation relation of the
bosonic fields. The commutation of the fields $\Phi $ and $\Theta $ (see the
Comment) was taken as $\left[ \Phi (x),\Theta (y)\right] =2\pi i\left[ \text{%
sgn}\left( x-y\right) -1\right] $ \cite{OFb}; $\Phi $ and $\Theta $ are
proportional to $\phi $ and $\widetilde{\phi }$ respectively in the
notations of the Letter. This regularization gives the correct
anticommutation relations between the right and left moving fermion
operators. The minus sign in front of the term $Z_{odd}$ in Eq.~(7) of our
Letter \cite{OFa} is a result of these anticommutation relations.

(ii) The authors of the Comment got the following expression for the Green's
function determining the tunneling DOS: $G^{\left( R\right) }\propto t^{-%
\frac 1{2g}}\left\langle \uparrow \right| e^{iH_{\uparrow
}t}e^{-iH_{\downarrow }t}\left| \uparrow \right\rangle $, where $H_{\uparrow
,\downarrow }=H_0\pm \frac{V_{bs}}{\pi \alpha }\cos \left[ \sqrt{g}\Phi
(x=0)\right] $. FG accept our conclusion that the $\Phi $ field at the
impurity site develops a finite average value, the fluctuations around which
are massive. In this case the analogy with the $X$-ray edge problem does not
hold and $\left\langle \uparrow \right| e^{iH_{\uparrow
}t}e^{-iH_{\downarrow }t}\left| \uparrow \right\rangle \propto $ $\exp
\left[ -2i\frac{V_{bs}}{\pi \alpha }\left\langle \uparrow \right| \cos
\left[ \sqrt{g}\Phi (x=0)\right] \left| \uparrow \right\rangle t\right] $.
Since the fluctuations of $\Phi (x=0)$ are massive, this correlator does not
have a power law prefactor. This leads to the tunneling DOS obtained in Ref. 
\cite{OFa}: 
\begin{equation}
\rho \left( \omega \right) \sim \omega ^{\frac 1{2g}-1},  \label{tdos}
\end{equation}
where $g<1$ for a repulsive interaction.

We show now that the different behavior of $\rho \left( \omega \right) $
obtained in the Comment is a result of a substitution of the backward
scattering problem by another problem, which is not equivalent to the
original one. In the calculation of $\left\langle \uparrow \right|
e^{iH_{\uparrow }t}e^{-iH_{\downarrow }t}\left| \uparrow \right\rangle $ FG
relied on their results for a finite chain of length $L$ with {\em open
boundary conditions}: $\psi (0)=\psi (L)=0$, where $\psi $ is the fermion
field \cite{FG}. This problem is equivalent to a ring that has a cut at $x=0$%
. FG used the operator $U=e^{i\pi J/2}$ to transform $H_{\downarrow }$ into $%
H_{\uparrow }$, where $J=\int_0^L\left( \rho _R(x)-\rho _L(x)\right) dx$.
This operation shifts $\Phi (x=0)\rightarrow \Phi (x=0)+\pi /\sqrt{g}$, but
should not influence $H_0$. One can check, however, that $UH_0U^{\dagger }$
gives an additional term proportional to $\rho _R(0)+\rho _L(0)-\rho
_R(L)-\rho _L(L)$ in the transformed Hamiltonian . Contrary to the problem
with open boundary conditions there are no reasons for this term to vanish
in the problem of a weak backward scattering. In the latter problem, as we
have shown \cite{OFa,OFb}, the field $\Phi (x=0,t)$ is frozen at low
temperatures. This leads to a vanishing of the current fluctuations
(proportional to $d\Phi (x=0,t)/dt$), but not of $\rho _R(0)+\rho
_L(0)\propto d\Phi (x=0,t)/dx$. This is why the low energy physics of the
weak backward scattering defect is not equivalent to cutting the wire.

We would like to emphasize that the vanishing of the dc conductance with a
power law behavior at low temperatures \cite{KF}, is not in contradiction
with the enhancement of the tunneling DOS. Since the current fluctuations
are suppressed at low energies, one should include instantons in the
calculation of the current--current correlation function \cite
{Larkin78,Schmid83} in order to obtain a non zero conductance. These
instantons correspond to jumps of $\Phi (x=0)$ between subsequent minima of
the term $\cos (\sqrt{g}\Phi (x=0))$ at some time. The configurations
without instantons do not contribute to the zero frequency conductance, but
they give the dominant contribution to the tunneling DOS. The first non
vanishing contribution to the conductance is due to two instanton
configurations. Their contribution to the tunneling DOS is much smaller than
the contribution of the configurations without instantons. Thus, the
description of the strong coupling regime by pinning of the $\Phi $ field,
which leads to a freezing of the current fluctuations, is consistent with
both the vanishing of the conductance and the enhancement of the tunneling
DOS.

We demonstrate that the result of the Letter follows from the Comment when $
\left\langle \uparrow \right| e^{iH_{\uparrow
}t}e^{-iH_{\downarrow }t}\left| \uparrow \right\rangle$
is calculated correctly. This clearly indicates that the question
about the anticommutation relations was raised by FG without serious grounds.

This work was supported by the Israel Academy of Science, Grant No.
801/94-1. \\

\noindent
Yuval Oreg and Alexander M. Finkel'stein$^{\dag}$ \\The Weizmann Institute of Science %
\\Rehovot 76100, Israel\\
$^{\dag}$ Also at the Landau Institute for Theoretical Physics, Russia.

\end{multicols}


\begin{references}
\bibitem{Comment}  M. Fabrizio and A.O. Gogolin, Phys. Rev. Lett. {\bf 78}, 4527 (1997)

\bibitem{OFa}  Y. Oreg and A.M. Finkel'stein, Phys. Rev. Lett. {\bf 76},
4230 (1996).

\bibitem{OFb}  Y. Oreg and A.M. Finkel'stein, Phys. Rev. B {\bf 53}, 10928
(1996) (See Sec. IIB).

\bibitem{FG}  M. Fabrizio and A.O. Gogolin, Phys. Rev. B {\bf 51}, 17 827
(1995).

\bibitem{KF}  C.L. Kane and M.P.A. Fisher, Phys. Rev. Lett. {\bf 68}, 1220
(1992).

\bibitem{Larkin78}  A.~I.~Larkin and P.~A.~Lee Phys. Rev. B {\bf 17} 1596
(1978).

\bibitem{Schmid83}  A. Schmid, Phys. Rev. Lett. {\bf 51} 1506 (1983).
\end{references}
\end{document}